\documentstyle[aps,prl]{revtex}

\begin{document}
\draft
\twocolumn

\title{Quantum Diffusion of H/Ni(111)
through Monte Carlo Wave Function Formalism}

\author{S. C. Badescu$^{1,2}$, S. C. Ying$^{1}$, and T. 
Ala-Nissila$^{1,2}$}

\address{$^{1}${Department of Physics, Box 1843,
Brown University, Providence, RI 02912--1843}\\
$^{2}${Helsinki Institute of Physics and
Laboratory of Physics, Helsinki University of Technology, P. O. Box 
1100,
FIN--02015 HUT, Espoo, Finland}}

\date{April 3, 2001}

\maketitle

\begin{abstract}

We consider a quantum system coupled to a dissipative background
with many degrees of freedom using the Monte Carlo Wave Function
method. Instead of dealing with a density matrix which can be very
high-dimensional, the method consists of integrating a stochastic
Schr\"{o}dinger equation with a non-hermitian damping term in the
evolution operator, and with random quantum jumps. The method is
applied to the diffusion of hydrogen on the Ni(111) surface below
100 K. We show that the recent experimental diffusion data for
this system can be understood through an interband activation process, followed by quantum tunnelling.
\end{abstract}


The study of a quantum system coupled to a large background
reservoir that leads to thermal fluctuations and dissipation in
the dynamical evolution of the system, is of central importance in
such fields as quantum optics \cite{ref1}, electronic conduction
in nano-structures \cite{ref2}, and diffusion of light adatoms on
surfaces \cite{ref3,ref4,ref4b}. The standard formalism for this problem 
is
through the master equation for the density matrix $\rho_S(t)$ of
the system \cite{ref3,ref4}.
However, this approach is not practical
for condensed matter systems such as a hydrogen adatom moving on a
metal surface. In this case, the density matrix would have
dimension $N^2$, where $N$ is the product of the number of sites 
considered on the
surface and the number of
vibrational states included at each site. Typically, $N$ would be
at least of the order of $10^4$ rendering a direct numerical
solution of the master equation unfeasible.

Recently, an alternative approach known as the Monte Carlo Wave
Function (MCWF) \cite{ref1} has been developed and applied to
solve these type of problems in the field of quantum optics. In
the MCWF approach, the evolution of a quantum state $|\Psi(t)\rangle$ is
described by a stochastic wave equation, in which the
original adiabatic
Hamiltonian $H_S$ is only a part of the evolution operator:
\begin{eqnarray}
|\Psi(t+\delta t)\rangle &=& \frac{f_0}{\sqrt{1-\delta p}}
\exp(\frac{-iH\delta t}{\hbar})|\Psi(t)\rangle
\label{Schrod}  \\
&+& \sum_{\mu}\frac{f_{\mu}}
{\sqrt{\delta p_{\mu}/\delta t}}C_{\mu}|\Psi(t)\rangle. \nonumber
\end{eqnarray}
Here the effect of each operator $C_{\mu}$ acting on the quantum
system represents a collision with the reservoir degrees of freedom
that takes the system from one quantum state to another. The new
Hamiltonian $H$ is non-Hermitian, built from $H_S$
with an imaginary part added to account for dissipation:
\begin{equation}
    H=H_S-\frac{i\hbar}{2}\sum_{\mu}C_{\mu}^{+}C_{\mu}.  \label{ImHam}
\end{equation}
The stochastic nature of quantum evolution is described by the
quantities $f_0$ and ${f_{\mu}}$. They are random numbers
such that the mean value of $f_{\mu}$ is
related to the scattering probabilities
\begin{eqnarray}
  \delta p_{\mu}=\delta t
  \langle\Psi(t)|C^+_{\mu}C_{\mu}|\Psi(t)\rangle,  \label{Ev4}
\end{eqnarray}
with $\langle f_{\mu}\rangle$=$\delta p_{\mu}$, and
$\langle f_0 \rangle$=$1-\delta p$, where $\delta p=\sum{p_{\mu}}$
gives the probability for coherent propagation under $H$. With
this choice of dynamics, it can be shown \cite{ref1} that the
quantity $\bar{\sigma}(t)$ obtained by averaging
$\sigma(t)=|\Psi(t)\rangle\langle \Psi(t)|$ over all possible
outcomes at time $t$ of the MCWF evolution equation, coincides
with the density matrix $\rho_S(t)$ obtained from the solution of
the so-called Lindblad form of the master equation \cite{ref5}:
\begin{eqnarray}
\dot{\rho_S}&=&\frac{i}{\hbar}[\rho_S,H_S]-\\
&\sum_{\mu}& \frac{1}{2}(C_{\mu}^{+} C_{\mu}\rho_S+
\rho_S C_{\mu}^{+} C_{\mu}-2C_{\mu}\rho_S 
C_{\mu}^{+})\label{MasEq}\nonumber
\end{eqnarray}
The equality between $\bar{\sigma}$ and $\rho_S$ holds at {\it all}
times $t$, provided that it holds at $t=0$. The
particular form of the collision operators chosen in Eq.
(\ref{Schrod}) is the most general one that preserves the
normalization and positive definiteness of the corresponding
$\rho_S(t)$.

It is the purpose of the present Letter to demonstrate how the
MCWF method can be used to tackle important transport problems in
condensed matter physics in cases where the number of degrees of
freedom is large enough ($N \gtrsim 10^4$) to make the density
matrix approach unfeasible. We consider here the case of a light
adatom moving on a metal surface under conditions where classical
activated hopping rate between potential wells is negligible
compared with the corresponding tunnelling rate. At present, there
does not exist a clear consensus on the details of the crossover
from the classical activated behavior to the quantum tunnelling
regime. In the Field Emission Microscopy (FEM) study \cite{ref6}
for Ni and W substrates and in the latest STM study  for H/Cu(001)
\cite{ref7}, a sharp crossover from classical diffusion to very
weak temperature dependence of diffusion was observed at a
temperature in the range of $60-100$ K. However, the Quasielastic
Helium Atom Scattering study for H/Pt(111) \cite{ref8} yields no
crossover down to $T \approx 100K$. For the H/Ni(111) system,
recent optical studies \cite{ref9} showed a crossover behavior
from the classical regime to a {\it second activated regime} with
a lower activation energy below $T \approx 100K$. This is in
contradiction with the FEM data on the same system, which showed a
crossover to a temperature independent diffusion at low
temperatures \cite{ref6}. Thus, while there is strong evidence
that diffusion proceeds through quantum tunnelling at low
temperatures, the detailed mechanisms for hydrogen diffusion on
different substrates are not yet understood. Previous theoretical
works do suggest that the details of the crossover is sensitive to
the shape  of the adsorption potential and not just determined by
the barrier alone \cite{ref3,ref4,ref4b}.

We will apply here the MCWF method to study the dynamics of
H/Ni(111). The low temperature activated behavior with a barrier
of about $90$ meV has been attributed to small polaron type
activated tunnelling \cite{ref9}. In our view, this is a highly
implausible explanation. First, the polaron
activation energy for H/Cu(001) \cite{ref7} was determined to be $\sim
3$meV, then the relaxation energy due to the adatom for
H/Ni(001) has been calculated to be 2.72meV \cite{ref4}, and in our recent calculations for H/Pt(111)
\cite{ref9b} we also find a relaxation energy of just a
few meV; the polaron activation
energy is a fraction of the relaxation energy
\cite{ref15}. We will show instead that the data can be explained in terms of
tunnelling from the first excited vibrational states of the H
adatom.

We construct a semi-empirical potential $U({\bf r})$ based on
available data as follows. The lowest energy adsorption sites are
assumed to be the {\it fcc} sites forming a 2D triangular lattice
$\{\bf{l}\}$ \cite{ref11} with a lattice constant $a = 2.581$
{\AA} (see Fig. 1). Also, the neighboring {\it hcp} sites at a
distance of $s=1.49$ {\AA} \cite{ref9} are taken to be equal in
energy \cite{ref9} (this is also supported by a recent {\it ab
initio} calculation \cite{ref12}). Second, we fix the barrier
between the {\it fcc} and {\it hcp} sites close to the value of
196 meV found in experiments \cite{ref9}. We use the vibrational
excitation energy of $94$ meV known from
\cite{ref12b},\cite{ref13}. $U({\bf r})$ is constructed from
localized Gaussians at both the {\it fcc} and {\it hcp} sites and
adjust the Gaussian parameters, obtaining a fitting with a band
gap  between the centers of the $A_0 \oplus A_1$ and the $E_0
\oplus  E_1$ bands of $\Delta = 96$ meV and a separation between
the lowest band and the top of the barrier between {\it fcc} and
{\it hcp} sites of $207$ meV.

The adiabatic Hamiltonian $H_S$ for our model is characterized by Bloch
states $\{|{\bf k},m\rangle \}$ with corresponding energy
$\{ \epsilon_{{\bf k},m}\}$. Here $m$ is the band index and ${\bf k}$ 
the
2D wave vector. The center positions and the bandwidths
for the first few bands are listed in Table I. The first two
branches form 1D representations ($A_0$ and $A_1$) of
the symmetry group of the 2D triangular lattice, while the next four
form 2D representations ($E_0$ and $E_1$).

We describe the H adatom as a linear superposition of energy 
eigenstates:
\begin{equation}
|\Psi(t)\rangle=
\sum_{m,{\bf k}}b_{{\bf k},m}(t)|{\bf k},m\rangle, \label{Prop1}
\end{equation}
with $\sum_{m,{\bf k}}|b_{{\bf k},m}|^2=1$. The frictional coupling to
the substrate through electronic and phononic excitations is modelled by
a general collision operator $C_{\mu}$ (\ref{Schrod}), through which we 
model {\it both} intraband {\it and} interband transitions.
It is represented as
\begin{equation}
C_{m_1,m_2,\bf{q}}=\Gamma_{m_1m_2,\bf{q}}^{1/2}\sum_{\bf{k}}|{\bf k}
+{\bf q},m_1\rangle \langle {\bf k},m_2|, \label{Op1}
\end{equation}
where $\Gamma$ is a (yet unspecified) transition rate, and
$\mu$ in Eq. (1) now becomes a multiple index with two band indices,
$\mu$=$(m1,m2,{\bf q})$.
Thus the probabilities for scattering $\delta p_{\mu}$ are given by
\begin{equation}
\delta p_{\mu}=\langle \Psi (t)|C^+_{\mu}C_{\mu}|\Psi(t)
\rangle \delta t=
\sum_{\bf{k}}|b_{{\bf k},m_2}|^2 \Gamma_{\mu} \delta t.
\end{equation}

An important feature of the model is that for the low energy bands of
interest, $A_0 \oplus A_1$ and $E_0 \oplus E_1$, the composite 
bandwidths
are much smaller than the energy gap $\Delta$ separating them (see Table 
I).
This means that we need to consider only two kinds of
transitions: interband transitions between the bands in the two
groups, and intraband transitions within each group. Since we do
not have microscopic expression for the scattering rates
$\Gamma_{\rm intra}$ and $\Gamma_{\rm inter}$ we
make one further simplification that
is $\Gamma_{\rm intra}$ = $\Gamma_{\rm inter} = \Gamma$.
Below, we will show
that the magnitude of $D$ is controlled by the parameter
$\gamma=\hbar\Gamma/\Delta_E$, where $\Delta_E$ is the width of the 
upper
composite band defined above.

In our numerical calculations, the substrate is represented by a 2D
hexagonal box consisting of $180 \times 180$ unit cells,
with fully periodic boundary
conditions. The size of the system is chosen such that the H
adatom does not spread outside the boundary during the observation
time $t$. To calculate the spatial $\alpha\beta$ elements of the
tracer diffusion coefficient of H, we used the expression
\begin{equation}
D_{\alpha \beta}(t)= \lim_{t\to\infty}\frac{1}{2t}
\langle(\hat{x}_{\alpha}-\langle\hat{x}_{\alpha} \rangle_0)
(\hat{x}_{\beta}-\langle\hat{x}_{\beta}\rangle_0)\rangle,
\label{DiffCoef}
\end{equation}
where $\hat{x}$ is the position operator. The average 
$\langle...\rangle$ in
Eq. (\ref{DiffCoef}) represents both the quantum
mechanical average in a given state as well as the ensemble
average over different initial states.
Statistical averages to compute $D$ were performed with $1500-6000$
initial states, for time intervals containing
up to $10^5$ collisions. With a code parallelized on $4$ processors, one 
point on the
Arrhenius plot takes $2-4$hrs, depending on the collision rates.

The symmetry of the lattice implies that the diffusion tensor
$D_{\alpha\beta}$ is diagonal. Fig. 2 shows the temperature
dependence of $D$ for $\gamma=1, 5,$ and $10$ on an Arrhenius
plot. There is clear activated behavior $D \propto e^{-E_a/k_BT}$,
with an activation energy $E_a = 98.1 \pm 0.5$ meV. This is in
excellent agreement with the experimental data of Cao {\it et al.}
\cite{ref9} shown in Fig. 2 as well, in the temperature regime
below 100 K where $E_a \simeq 105$ meV. Obviously, with the
inclusion of only the lowest bands in the present calculation, we
cannot account for the classical high temperature region above 100
K where $E_a \simeq 196$ meV \cite{ref9}. We can give a good
qualitative description of the quantum regime, though, where the
numerical results above indicate that the observed Arrhenius
behavior for $D$ corresponds to {\it activated quantum
tunnelling}.

The result for the temperature dependence can be understood from
the values of the bandwidths listed in Table 1. The bandwidths of
the $\{E_0,E_1\}$ states are more than one order of magnitude
larger than for the lower bands (the delocalization was observed
also in a recent experiment \cite{ref12b}). Thus, diffusion
proceeds mainly via a collisions excitation to the upper band,
followed by tunnelling to neighboring sites and de-excitation to
the lower bands again. It is the Bose-Einstein factor $n(\omega)$
($\hbar\omega=\Delta$), needed to ensure detailed balance in
thermal equilibrium \cite{ref14}, that leads naturally to the
activated Arrhenius behavior with an activation energy close to
the energy gap $\Delta$. Although the Arrhenius behavior of $D$
does not depend on the ratio $\gamma$, its absolute magnitude is
best fit to the experimental data by choosing $\gamma \approx 10$.
This should be taken only as an effective ratio between tunnelling
and scattering, because {\it e.g.} polaron effects
\cite{ref16,ref17} which lead to a broadening of the levels and a
reduction in the tunnelling rate have been left out in the present
calculation.

The MCWF methods gives insight into the quantum dynamics by
allowing to follow the dynamics of wave packets in real space and
time. In Fig. 1 we show two typical trajectories, tracing the
evolution of $\langle \hat{\bf r} \rangle$ for a wave packet. The
larger length scale for the trajectory at 110K reflects the larger
value of the diffusion coefficient, which is due to a higher
excitation rate into the upper bands. The trajectory at 70 K has
points where the particle is in the ground state for a longer time
and, by comparison to the trajectory at 110 K, it has less
coherent propagation intervals in the upper band. The other point
to note is that there are coherent propagation regions with
tunnelling through several sites before a de-excitation. This can
be quantified by studying the tunnelling length distribution
$P_{\ell}$.
 We define the tunnelling length $\ell$ as
the distance travelled by a wave packet in the upper band before
it suffers a collision.  It is found that asymptotically
$P_{\ell}$ decreases exponentially with $\ell$, while it obeys a
Poisson-like distribution at small values of $\ell$ ($\ell\leq
s$).
This is similar to the jump distribution in the  classical regime
\cite{ref18}. Regarding the dependence of $D$ on $\gamma$, we have
done
 simulations at $T=70$ K and $T=110$ K in the range $0.1
\leq \gamma \leq 10$ and found that $D \propto \gamma^{-1}$ in this
range.
This inverse power law dependence on $\gamma$ is similar to the
dependence of $D$ on the microscopic friction $\eta$ in the
classical regime \cite{ref19,ref19b}. However, the influence of the
geometrical factor on the dependence of the jump distribution on
$\gamma$ seems rather different from the classical case. The
crossover of the dependence on $\gamma$ or $\eta$ from the quantum
to classical behavior is a subject worthy of further
investigations.

To summarize, we have demonstrated through a model study of H
diffusion on Ni(111) that the MCWF method is a powerful tool in
the study of quantum transport problems with many degrees of
freedom. In addition, the real space nature of the method allows
one to extract interesting information about the dynamics of wave
functions, not easily available in other means. As opposed to the
small polaron mechanism suggested earlier \cite{ref9}, our results
suggest that the low temperature diffusion behavior observed in
the work of Cao {\it et al.} for H/Ni(111) \cite{ref9} has its
origin in the tunnelling of the hydrogen adatom from the first
vibrational excited state. We plan to apply the same MCWF
formalism to investigate other quantum diffusion systems, such as
H/Pt(111) \cite{ref8} and H/Cu(001) \cite{ref7}, which show
qualitatively different behaviors from H/Ni(111) \cite{ref9}. The
key is to start with a reliable adsorption potential through a
combination of first-principle calculation and empirical inputs.

Acknowledgements: This work has been in part
supported by the Academy of Finland
through its Center of Excellence program. We wish to than K.-A.
Suominen for introducing the MCWF method to us, and O. Trushin and
P. Salo for useful discussions.

\newpage

\begin{table}

\caption{Bandwidths $\Delta \epsilon_m$ and
band centers $\epsilon_m$ for branches $1-6$. Groups
$1-2$ and $3-6$ form the composite bands $A_0\oplus A_1$ and $E_0\oplus 
E_1$. }
    \begin{center}
         \begin{tabular}{lll}
          $m$ & $\Delta \epsilon_m$(meV) & $\epsilon_m$(meV) \\ \hline
          $1 \,\,(A_0)$    & 0.008 & 104.487  \\ \hline
          $2 \,\,(A_1)$  &  0.008 & 104.497  \\ \hline
          $3 \, \, (E_0)$  &  0.017 & 200.346  \\
          $6 \, \, $      & 0.017 & 200.721  \\ \hline
          $4 \, \, (E_1)$  &  0.146 & 200.446  \\
          $5 \, \, $  &  0.146 & 200.621
         \end{tabular}
     \end{center}
   \end{table}


\begin{figure}[htbp]
\caption{
(a) Trajectories at $T=70$ K (smaller set) and $110$ K (larger set). 
$\gamma=1$
and the observation time was
$3.1\times 10^{-2}s$.
(b) Details of the path at $T=70$ K. The black circles are excitations 
or de-excitations. Between two such
consecutive points there are usually several random changes of the
momentum}
\end{figure}

\begin{figure}[htbp]
\caption{Temperature dependence of $D$ between 80 K and 140 K, for
$\gamma=1, 5, 10$. The Arrhenius behavior is evident. The
experimental data of Cao {\it et al.} [9] are shown for
comparison. For $\gamma =10$, a prefactor $D_0$ of
$2.71\times10^{9}{\AA}^{2}/s$ is obtained. The experimental value
of the prefactor $D_0$ is  $2.4 \times 10^{9} {\AA}^{2}/s$ [9]}
\end{figure}

   \newpage

   \begin{figure}[htbp]
   \unitlength1cm
    \begin{center}
     \begin{picture}(8,6.5)
       \includegraphics{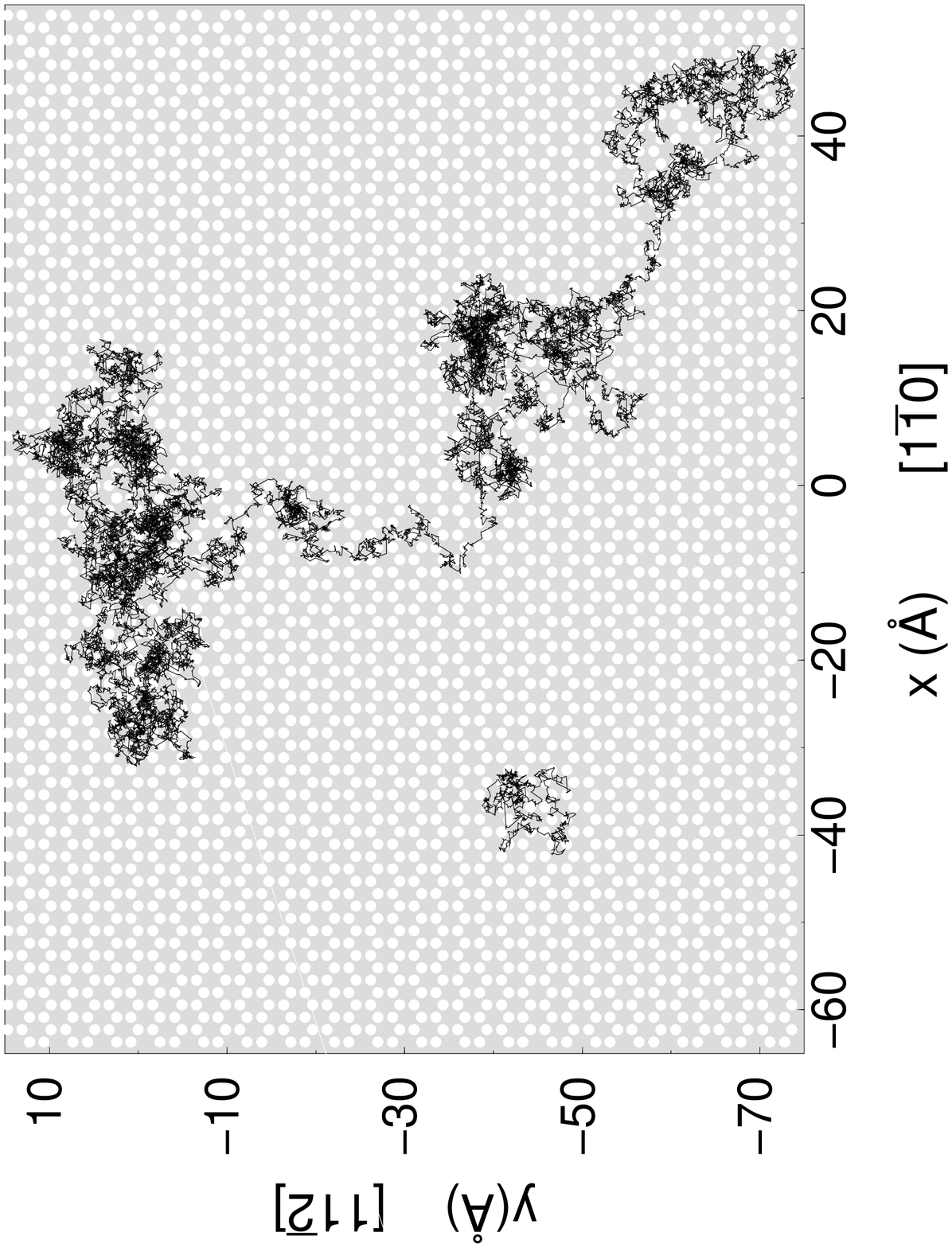}
    (a)
     \end{picture}

     \begin{picture}(8,6.5)
       \includegraphics{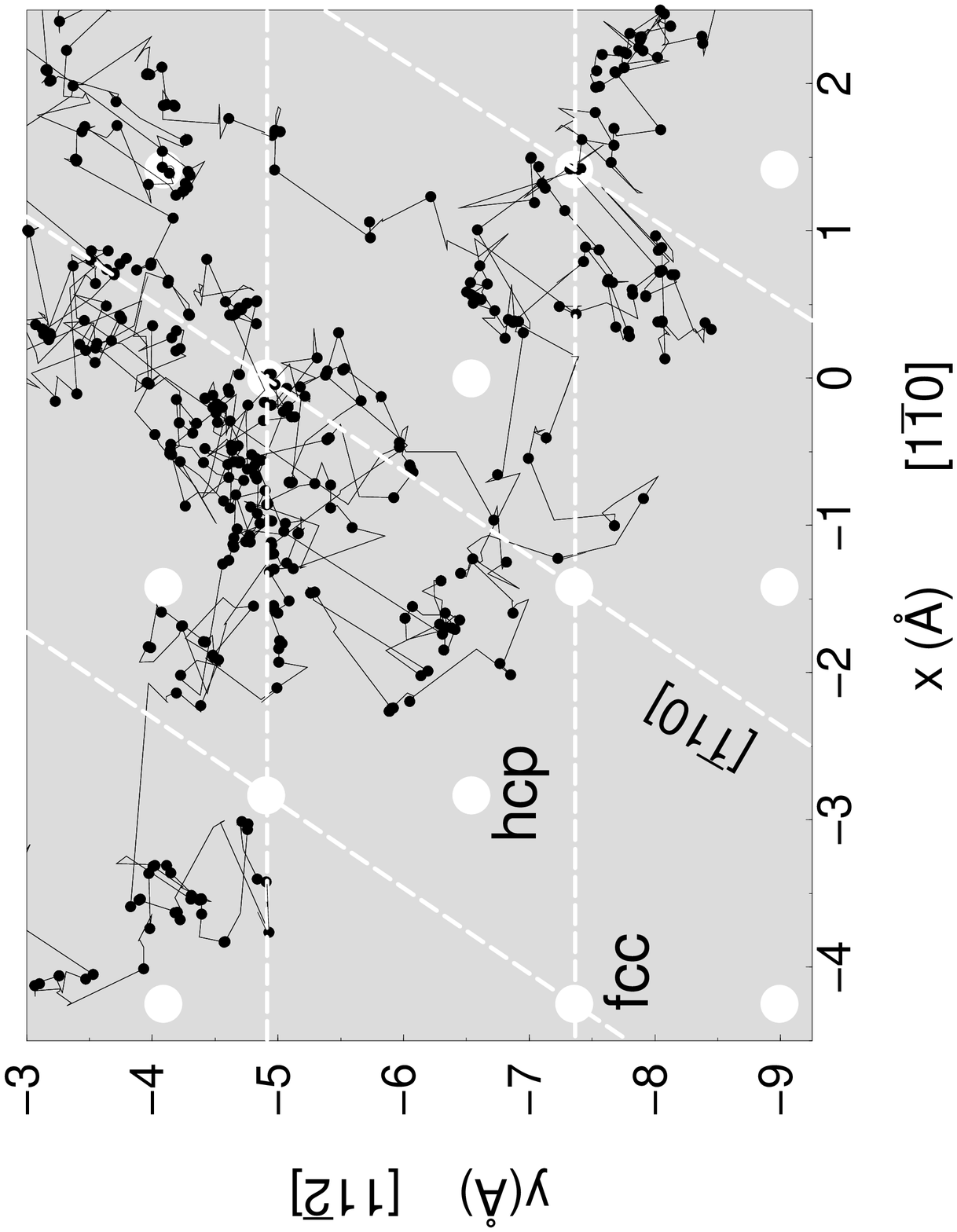}
    (b)
     \end{picture}
    \end{center}
    \end{figure}

   \begin{figure}[htbp]
   \unitlength1cm
    \begin{center}
     \begin{picture}(8,6.5)
       \includegraphics{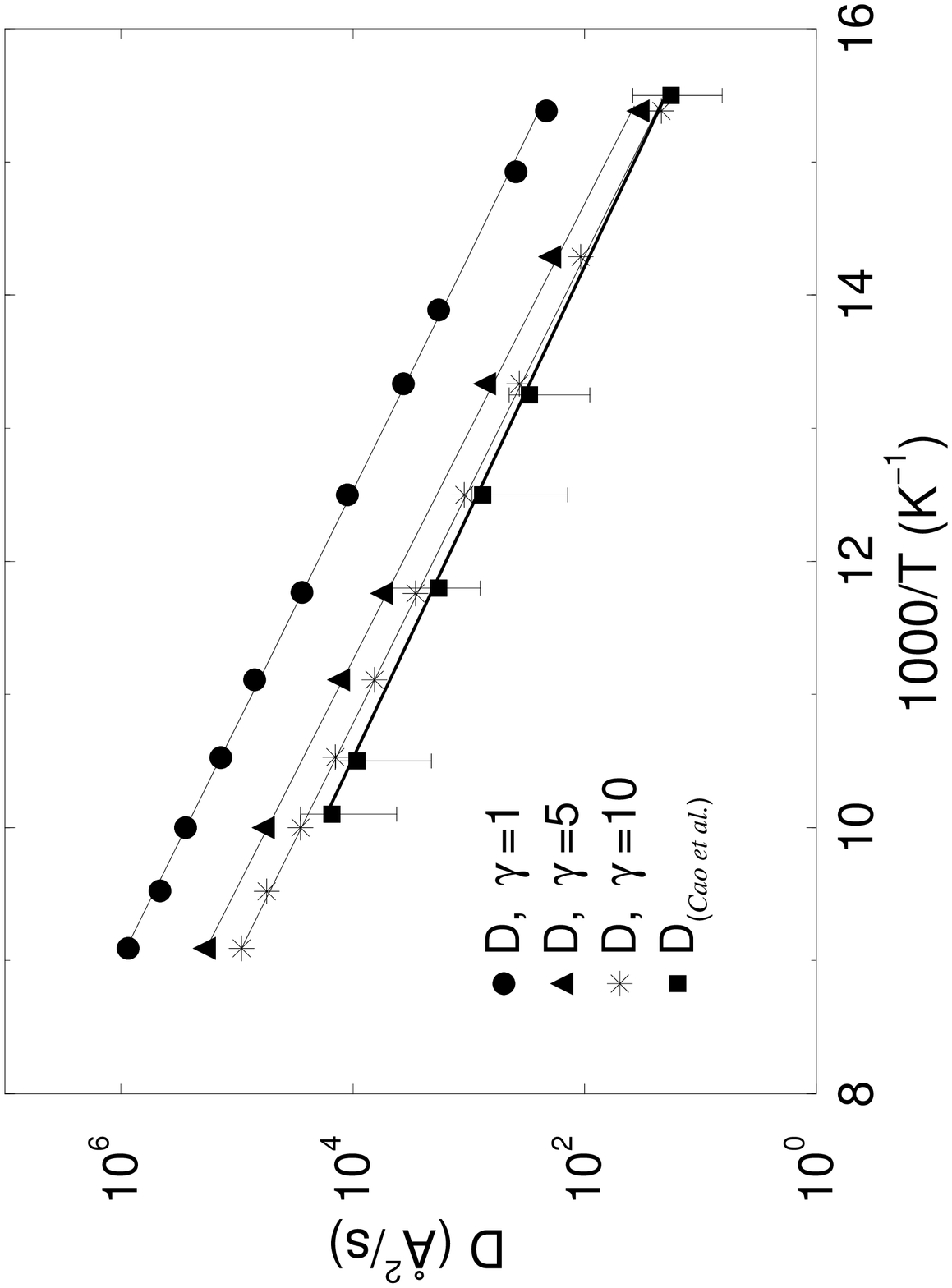}
     \end{picture}
    \end{center}
   \end{figure}


\begin{thebibliography}{200}
\bibitem{ref11}
The present assumption of a 2D model should be well justified
based on the recent evidence that there is no significant
mixing between the vertical
and horizontal modes, due to the very
different characteristic freqencies \cite{ref12b}.
\end{thebibliography}
\end{document}